\documentclass[conference]{IEEEtran}
\IEEEoverridecommandlockouts
\usepackage{cite}
\usepackage{amsmath,amssymb,amsfonts}
\usepackage{graphicx}
\usepackage{textcomp}
\usepackage[dvipsnames]{xcolor}
\def\BibTeX{{\rm B\kern-.05em{\sc i\kern-.025em b}\kern-.08em
  T\kern-.1667em\lower.7ex\hbox{E}\kern-.125emX}}

\usepackage{caption}
\usepackage{makecell}

\usepackage{multirow}
\usepackage{hhline}
\DeclareMathOperator*{\argmax}{arg\,max}
\newtheorem{definition}{Definition}

\usepackage{algorithm}
\usepackage{algpseudocode}
\usepackage{algpascal}
\newlength\myindent
\usepackage{xpatch}

\makeatletter
\xpatchcmd{\algorithmic}
 {\ALG@tlm\z@}{\leftmargin\z@\ALG@tlm\z@}
 {}{}
\makeatother

\usepackage{rsfso}

\usepackage{pifont}
\usepackage{float}

\usepackage{siunitx}

\usepackage{tikz}
\usepackage{pgfplots}
\pgfplotsset{width=7cm,compat=1.3}

\usepackage{makecell}

\begin{document}

\title{Effects of Differential Privacy and Data Skewness\\ on Membership Inference Vulnerability\\
}

\author{\IEEEauthorblockN{Stacey Truex, Ling Liu, Mehmet Emre Gursoy, Wenqi Wei, Lei Yu}
\IEEEauthorblockA{School of Computer Science, Georgia Institute of Technology, Atlanta, Georgia, USA}
}

\maketitle

\begin{abstract}
Membership inference attacks seek to infer the membership of individual training instances of a privately trained model. This paper presents a membership privacy analysis and evaluation system, called MPLens, with three unique contributions. First, through MPLens, we demonstrate how membership inference attack methods can be leveraged in adversarial machine learning. Second, through MPLens, we highlight how the vulnerability of pre-trained models under membership inference attack is not uniform across all classes, particularly when the training data itself is skewed. We show that risk from membership inference attacks is routinely increased when models use skewed training data. Finally, we investigate the effectiveness of differential privacy as a mitigation technique against membership inference attacks. We discuss the trade-offs of implementing such a mitigation strategy with respect to the model complexity, the learning task complexity, the dataset complexity and the privacy parameter settings. Our empirical results reveal that (1) minority groups within skewed datasets display increased risk for membership inference and (2) differential privacy presents many challenging trade-offs as a mitigation technique to membership inference risk. 
\end{abstract}


\section{Introduction}
\setlength{\textfloatsep}{0pt}
Machine learning-as-a-service (MLaaS) has seen an explosion of interest with the development of cloud platform services. Many cloud service providers, such as Amazon~\cite{amazonml}, Google~\cite{googleml}, IBM~\cite{ibmml}, and Microsoft~\cite{Copeland:2015:MAP:2840185}, have launched such MLaaS platforms. These services allow consumers and application companies to leverage powerful machine learning and artificial intelligence technologies without requiring in-house domain expertise. Most MLaaS platforms offer two categories of services. (1) Machine learning model training. This type of service allows users and application companies to upload their datasets (often sensitive) and perform task-specific analysis including private machine learning and data analytics. The ultimate goal of this service is to construct one or more trained predictive models. (2) Hosting service for pre-trained models. This service provides pre-trained models with a prediction API. Consumers are able to select and query such APIs to obtain task specific data analytic results on their own query data. 

With the exponential growth of digital data in governments, enterprises, and social media, there has also been a growing demand for data privacy protections, leading to legislation such as HIPAA~\cite{act1996health}, GDPR~\cite{regulation2016regulation}, and the 2018 California privacy law~\cite{ca2018bill}. Such legislation puts limits on the sharing and transmission of the data analyzed by these platforms and used to train predictive models. All MLaaS providers and platforms are therefore subject to the compliance of such privacy regulations.

With the new opportunities of MLaaS and the growing attention on privacy compliance, we have seen a rapid increase in the study of potential vulnerabilities involved in deploying MLaaS platforms and services. Membership inference attacks and adversarial examples represent two specific vulnerabilities against deep learning models trained and deployed using MLaaS platforms. With more mission critical cyber applications and systems using machine learning algorithms as a critical functional component, such vulnerabilities are a major and growing threat to the safety of cyber-systems in general and the trust and accountability of algorithmic decision making in particular.

\textbf{Membership Inference.}
Membership inference refers to the ability of an attacker to infer the membership of training examples used during model training. We call a membership inference a black-box attack if the attacker only has the access to the prediction API of a privately trained model hosted by a MLaaS provider. A black-box attacker therefore does not have any knowledge of either the private training process or the privately trained model. 

Consider a financial institution with a large database of previous loan applications. This financial institution leverages its large quantity of data in conjunction with a MLaaS platform to develop a predictive model. The goal of constructing such a model is to provide prediction when given individuals' personal financial data as input, such as a likelihood of being approved for a loan under multiple credit evaluation categories. This privately trained predictive model is then deployed with a MLaaS API at different offices of this financial institution. When potential applicants provide their own data to the query API, they receive some prediction statistics on their chance of receiving the desired loan and perhaps even the ways in which they may improve their likelihood of loan approval. In the context of the loan application approval model, a membership inference attack considers a scenario wherein a user of the prediction service is an attacker. This attacker provides data of a target individual $X$ and, based on the model's output, the attacker tries to infer if $X$ had applied for a loan at the given financial institution.

Both the financial institution and individual $X$ have an interest in protecting against such membership inference attacks. Loan applicants (such as $X$) consider their applications and financial data to be sensitive and private. They do not want their loan to be public knowledge. The financial institution also owns the training dataset and the privately trained model and likely considers their training data to not only be confidential for consumer privacy but also an organizational asset~\cite{lake2013data}. It is therefore a high priority for MLaaS providers and companies to protect their private training datasets against membership inference risks for maintaining and increasing their competitive edge.

\textbf{Adversarial Machine Learning.}
The second category of vulnerability is the adversarial input attacks against modern machine learning models, also referred to as adversarial machine learning (ML). Adversarial ML attacks are broadly classified into two categories: (1) evasion attacks wherein attackers aim to mislead pre-trained models and cause inaccurate output; and (2) poisoning attacks which aim to generate a poisonous trained model by manipulating its construction during the training process. Such poisoned models will misbehave at prediction time and can be deceived by the attacker. Adversarial deep learning research to date has been primarily centered on the generation of adversarial examples by injecting the minimal amount of perturbation to benign examples required to either (1) cause a pre-trained classification model to misclassify the query examples with high confidence or (2) cause a training algorithm to produce a learning model with inaccurate or toxic behavior. 

The risks of adversarial ML have triggered a flurry of attention and research efforts on developing defense methods against such deception attacks. As predictive models take a critical role in many sensitive or mission critical systems and application domains, such as healthcare, self-driving cars, and cyber manufacturing, there is a growing demand for privacy preserving machine learning which is secure against these attacks. For example, the ability of adversarial ML attacks to trick a self-driving car into identifying a stop sign as a speed limit sign poses a significant safety risk. Given that most of the prediction models targeted are hosted by MLaaS providers and kept private with only a black-box access API, one common approach in developing adversarial example attacks is to use a substitute model of the target prediction model. This substitute model can be constructed in two steps. First, use membership inference methods to infer the training data of the target model and its distribution. Then, utilize this data and distribution information to train a substitute model. 

Interestingly, most of the research efforts to date have been centered on defense methods against already developed adversarial examples~\cite{liu2019deep} but few efforts have been dedicated to the countermeasures against membership inference risks and an attacker's ability to develop an effective substitute model. 

In this paper, we focus on investigating two key problems regarding model vulnerability to membership inference attacks. First, we are interested in understanding how skewness in training data may impact the membership inference threat. Second, we are interested in understanding a frequently asked question: can differentially private model training mitigate membership inference vulnerability? This includes several related questions, such as when such mitigation might be effective and the reasons why differential privacy may not always be the magic bullet to fully conquer all membership inference threats.

\textbf{Machine learning with Differential Privacy.}
Differential privacy provides a formal mathematical framework, which bounds the impact of individual instances on the output of a function when this function is constructed in a differentially private manner. In the context of deep learning, a deep neural network model is said to be differentially private if its training function is differentially private therefore guaranteeing the privacy of its training data. Thus, conceptually, differential privacy provides a natural mitigation strategy against membership inference threats. If training processes could limit the impact that any single individual instance may have on the model output, then the differential privacy theory~\cite{dwork2008differential} would guarantee that an attacker would be incapable of identifying with high confidence that an individual example is included in the training dataset. Additionally, recent research has indicated that differential privacy also has a connection to the model robustness against adversarial examples~\cite{lecuyer2019certified}.

Unfortunately, differential privacy can be challenging to implement efficiently in deep neural network training for a number of reasons. First, it introduces a substantial number of parameters into the machine learning process, which already has an overwhelming number of hyper-parameters for performance tuning. Second, existing differentially private deep learning methods tend to have a high cost in prolonged training time and lower training and testing accuracy. The effort for improving deep learning with differential privacy has therefore been centered on improving training efficiency and maintaining high training accuracy~\cite{abadi2016deep, yu2019differentially}. We argue that balancing privacy, security, and utility remains an open challenge for supporting differential privacy in the context of machine learning in general, and deep neural network model training in particular. 

{\bf Contributions of the paper.\/}
In this paper, we present a privacy analysis and compliance evaluation system, called MPLens, which investigates Membership Privacy through a multi-dimensional Lens. MPLens aims to expose membership inference vulnerabilities, including those unique to varying distributions of the training data. We also leverage MPLens to investigate differential privacy as a mitigation technique for membership inference risk in the context of deep neural network model training. Our privacy analysis system can serve for both MLaaS providers and data scientists to conduct privacy analysis and privacy compliance evaluation. This paper presents our initial design and implementation of MPLens and it makes three original contributions. 

First, through MPLens, we demonstrate how membership inference attack methodologies can be leveraged in adversarial ML. Datasets developed by using the model prediction API for MLaaS not only reveal private information about the training dataset, such as the underlying distributions of the private training data, but they can also be used in developing and validating the adverse utility of adversarial examples.

Second, MPLens identifies and highlights that the vulnerability of pre-trained models to the membership inference attack is not uniform when the training data itself is skewed. We show that risk from membership inference attacks is routinely increased when models use skewed training data. This vulnerability variation becomes particularly acute in federated learning environments wherein participants are likely to hold information representing different subsets of the population and therefore may incur different vulnerability to attack due to their participation.
We argue that an in-depth understanding of such disparities in privacy risks represent an important aspect for promoting fairness and accountability in machine learning.

Finally, we investigate the effectiveness of differential privacy as a mitigation technique against membership inference attacks, with a focus on deep learning models. We discuss the trade-offs of implementing such a mitigation strategy for preventing membership inference and the impact of differential privacy on different classes when deep neural network (DNN) models are trained using skewed training datasets.

\section{Membership Inference Attacks}

Attackers conducting membership inference attacks seek to identify whether or not an individual is a member of the dataset used to train a particular target machine learning model. We discuss the definition and the generation of membership inference attacks in this section, which will serve as the basic reference model of membership inference. 

\subsection{Attack Definition}

In studying membership inference attacks there are two primary sets of processes at play: (1) the training, deployment, and use of the machine learning model which the attacker is targeting for inference and (2) the development and use of the membership inference attack. Each of these two elements has guiding pre-defined objectives impacting respective outputs. 

\subsubsection{Machine Learning Model Training and Prediction}

The training of and prediction using the machine learning model which the attacker is targeting may be formalized as follows. Consider a dataset $D$ comprised of $n$ training instances $(x_1, y_1),(x_2, y_2), ...,(x_n, y_n)$ with each instance $x_i$ containing $m$ features, denoted by $x_i = (x_{i,1}, x_{i,2}, ..., x_{i,m})$, and a class value $y_i \in \mathbb{Z}_k$, where $k$ is a finite integer $\geq 2$. Let $F_t : \mathbb{R}^m \rightarrow \mathbb{R}^k$ be the target model trained using this dataset $D$. $F_t$ is then deployed as a service such that users can provide a \textit{feature vector} $\mathbf{x} \in \mathbb{R}^m$, and the service will then output a probability vector $\mathbf{p} \in \mathbb{R}^k$ of the form $\mathbf{p} = (p_1,p_2,...,p_k)$, where $p_i \in [0,1] \forall i$, and $\sum_{i=1}^k p_i = 1$. The prediction class label $y$ according to $F_t$ for a feature vector $\mathbf{x}$ is the class with highest probability value in $\mathbf{p}$. Therefore $y = \argmax_{i \in \mathbb{Z}_k} F_t(\mathbf{x})$. 

\subsubsection{Membership Inference Definition}

Given some level of access to the trained model $F_t$ the attacker conducts his or her own training to develop a binary classifier which serves as the membership inference attack model. The most limited access environment in which an attacker may conduct the membership inference attack is the black-box access environment. That is, an environment wherein the attacker may only query the target model $F_t$ through some machine learning as a service API and receive only the corresponding prediction vectors. 

Let us consider an attacker with such black-box access to $F_t$. Given only a query input $\mathbf{x}$ and output $F_t(\mathbf{x})$ from some target model $F_t$ trained using a dataset $D$, the membership inference attacker attempts to identify whether or not $\mathbf{x} \in D$. 

\begin{table}[ht]
  \centering
  \resizebox{\columnwidth}{!}{\begin{tabular}{|c|c|}
    \hline
    Dataset     & Accuracy of Membership Inference (\%) \\
    \hline
    \hline
    Adult           & 59.89                             \\
    MNIST           & 61.75                             \\
    CIFAR-10        & 90.44                             \\
    Purchases-10    & 82.29                             \\
    Purchases-20    & 88.98                             \\
    Purchases-50    & 93.71                             \\
    Purchases-100   & 95.74                             \\ 
    \hline
  \end{tabular}}\newline
  \caption{\small{Attack accuracies targeting decision tree models. Baseline accuracy against which to compare results is 50\%.}}
  \label{tab:mia_dt}
\end{table}

Many different datasets and model types have demonstrated vulnerability to membership inference attacks in black-box settings. Table \ref{tab:mia_dt} reports 5 accuracy results for black-box attackers targeting decision tree models for problems ranging from binary classification (Adult) to 100-class classification (Purchases-100). We note that all experiments evaluated the attack model against an equal number of instances in the target training dataset $D$ as those not in $D$. The baseline membership inference accuracy is therefore 50\%. We refer readers to~\cite{truex2019demystifying} for more details on these datasets and experimental set up. 

These results demonstrate both the viability of membership inference attacks as well as the variation in vulnerability between datasets. This accentuates the need for practitioners to evaluate their system's specific vulnerability.

Recently, researchers showed similar membership inference vulnerability in settings where attackers have white-box access to the target model, including the output from the intermediate layers of a pre-trained neural network model or the gradients for the target instance~\cite{nasr2018comprehensive}. Interestingly, this study showed that the intermediate layer outputs, in most cases, do not lead to significant improvements in attack accuracy. For example, with the CIFAR-100 dataset and AlexNet model structure, a black-box attack achieves 74.6\% accuracy while the white-box attack achieves 75.18\% accuracy. This result further supports the understanding that the attackers can gain sufficient knowledge from only the black-box access to the pre-trained models which is common in MLaaS platforms. Attackers do not require either full or even partial knowledge of the pre-trained target model as black-box attacks include the \textit{primary} source of membership inference vulnerability.

\subsection{Attack Generation}

\begin{figure}
  \centering
  \includegraphics[width=\columnwidth]{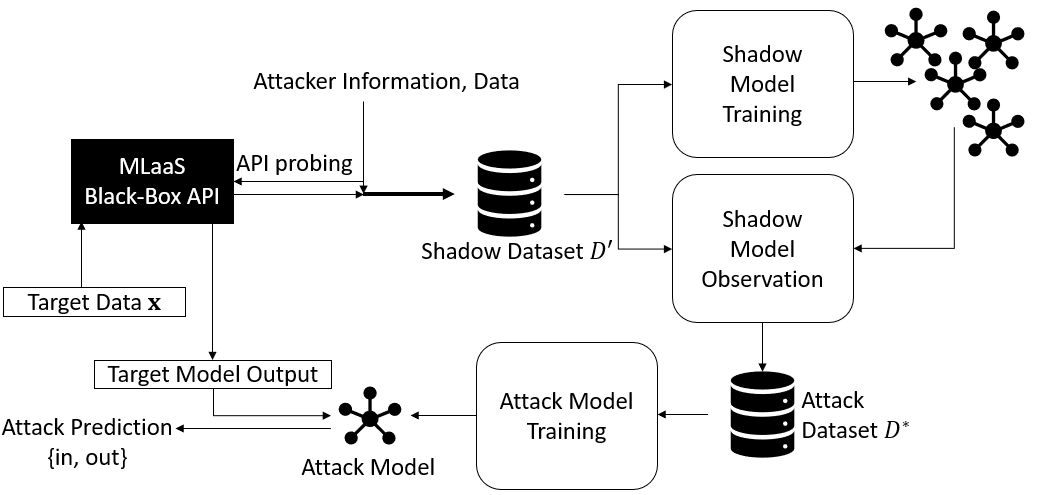}
  \caption{\small{Workflow of the membership inference attack generation.}}
  \label{fig:attackgeneration}
\end{figure}

The attack generation process can vary significantly based on the power of the attacker. For example, the attack proposed in~\cite{yeom2018privacy} requires knowledge of the training error of $F_t$. The attack technique proposed in~\cite{shokri2017membership}, however, requires computational power and involves the training of multiple machine learning models. The techniques proposed in~\cite{salem2018ml} are different still in that they require the attacker to develop effective threshold values. Figure~\ref{fig:attackgeneration} gives a workflow sketch of membership inference attack generation algorithm. We use the shadow model technique documented in~\cite{shokri2017membership} and~\cite{truex2019demystifying} to describe the attack generation process of membership inference attacks, while noting that many of the processes may be applicable to other attack generation techniques.

\subsubsection{Generating Shadow Data and Substitute Models}
In the shadow model technique, an attacker must first generate or access a \textit{shadow dataset}, a synthetic labeled dataset $D'$ to mirror the data in $D$. While~\cite{shokri2017membership} and~\cite{truex2019demystifying} both outline potential approaches to generating such a synthetic dataset from scratch, we would like to note that in many cases, attackers may also have examples of their own which can be used as seeds for the shadow data generation process or to bootstrap their shadow dataset. Consider our example of the financial institution. A competitor to the target institution may in fact have their own customer data, which could be leveraged to bootstrap a shadow dataset.

Once the attacker has developed the shadow dataset $D'$, the next phase of the membership inference attack is to leverage $D'$ to train and observe a series of \textit{shadow models}. Specifically, the shadow dataset is used to train multiple shadow models each of which is designed to emulate the behavior of the target model. Each shadow model is trained on a subset of the shadow dataset $D'$. As the attacker knows which portion of $D'$ was provided to each shadow model, the attacker may then observe the shadow models' behavior in response to instances which were in their training set versus behavior in response to those that were held out.

\subsubsection{Generating Attack Datasets and Models}

Attackers use the observations of the shadow models to develop an attack dataset $D^*$ which captures the difference between the output generated by the shadow models for instances included in the training data and those previously unseen by models.

Once the attack dataset $D^*$ has been developed, $D^*$ is used to generate a binary classifier $F_a$ which provides predictions on whether an instance was previously known to a model based on the model's output from that instance. At attack time this binary classifier may the be deployed against the target model service in a black-box setting. The attack model takes as input prediction vectors of the same structure as those provided by the shadow models and contained within $D^*$ and produces as output a prediction of $0$ or $1$ representing ``out'' and ``in'' respectively with the former indicating an instance that \textit{was not} in the training dataset of the target model and the latter indicating an instance that \textit{was} included.

The totality of these two phases: (1) generating shadow data and substitute models and (2) generating attack datasets and models, constitute the primary processes for constructing the membership inference attack. 

\begin{table*}[!t]
  \begin{tabular}{|c|c|c|c|c|c|c|c|c|}
    \hline
     \multirow{3}{*}{Degree of Noise} & \multicolumn{4}{c|}{Attack Accuracy (\%)} & \multicolumn{4}{c|}{Attack Accuracy (\%)} \\
     & \multicolumn{4}{c|}{with Noisy Target Data} & \multicolumn{4}{c|}{with Noisy Shadow Data} \\
     \cline{2-9}
                & CIFAR-10  & Purchases-10  & Purchases-20  & Purchases-50  & CIFAR-10  & Purchases-10  & Purchases-20  & Purchases-50  \\
     \hline
     0          & 67.49     & 66.69         & 80.70         & 88.52         & 67.49     & 66.69         & 80.70         & 88.52         \\ 
     0.1        & 65.37     & 68.72         & 80.40         & 86.38         & 66.85     & 67.37         & 80.23         & 88.81         \\ 
     0.2        & 63.88     & 66.01         & 77.47         & 85.85         & 65.36     & 66.86         & 81.20         & 88.35         \\ 
     0.3        & 60.43     & 62.90         & 73.93         & 84.66         & 64.74     & 67.46         & 80.32         & 88.84         \\ 
     0.4        & 60.48     & 60.07         & 68.23         & 83.21         & 62.64     & 66.91         & 80.14         & 88.63         \\ 
     0.5        & 58.33     & 58.29         & 64.73         & 79.12         & 60.94     & 67.61         & 80.36         & 88.17         \\ 
     0.6        & 57.53     & 57.58         & 61.51         & 73.50         & 60.09     & 66.68         & 80.08         & 88.54         \\ 
     0.7        & 55.97     & 54.94         & 59.78         & 70.43         & 58.92     & 67.67         & 80.83         & 88.58         \\ 
     0.8        & 55.35     & 54.44         & 58.21         & 67.16         & 58.66     & 66.73         & 80.49         & 88.54         \\ 
     0.9        & 54.07     & 54.03         & 57.91         & 65.21         & 57.57     & 68.06         & 80.52         & 87.84         \\ 
     1.0        & 53.95     & 52.72         & 56.02         & 62.44         & 56.55     & 67.32         & 80.43         & 87.70         \\  
     \hline
  \end{tabular}
  \caption{\small{Impact of noisy target data and noisy shadow data}}
  \label{tab:noisy_data}
  \vspace{-\baselineskip}
\end{table*}

\section{Characterization of Membership Inference}

\subsection{Impact of Model Based Factors on Membership Inference}

The most widely acknowledged factor impacting vulnerability to membership inference attacks is the degree of overfitting in the trained target model. Shokri et al. \cite{shokri2017membership} demonstrate that the more over-fitted a DNN model is, the more it leaks under membership inference attacks. Yeom et al.~\cite{yeom2018privacy} investigated the role of overfitting from both the theoretical and the experimental perspectives. While their results confirm that models become more vulnerable as they overfit more severely, the authors also state that overfitting is not the only factor leading to model vulnerability under the membership inference attack. Truex et al.~\cite{truex2019demystifying} further demonstrate that several other model based factors also play important roles in causing model vulnerability to membership inference, such as classification problem complexity, in-class standard deviation, and the type of machine learning model targeted. 

\subsection{Impact of Attacker Knowledge on Membership Inference}

Another category of factors that may cause model vulnerability to the membership inference attacks is the type and scope of knowledge which attackers may have about the target model and its training parameters. For example, Truex et al.~\cite{truex2019demystifying} identified the impact that attacker knowledge with respect to both the training data of the target model and the target data have on the accuracy of the membership inference attack. This was evaluated by varying the degree of noise in the shadow dataset and target data used by the attacker. Table~\ref{tab:noisy_data} shows the experimental results on four datasets with four types of learning tasks. The datasets include the CIFAR-10 dataset which contains 32$\times$32 color images of $10$ different classes of objects while the Purchases datasets were developed from the Kaggle Acquire Valued Shoppers Challenge dataset containing the shopping history of several thousand individuals. Each instance in the Purchases datasets then represents an individual and each feature represents a particular product. If an individual has a purchase history with this product in the Kaggle Acquired Valued Shoppers Challenge dataset, there will be a 1 for the feature and otherwise a 0. The instances are then clustered into different shopping profile types which are treated as the classes. Table~\ref{tab:noisy_data} reports results for Purchases datasets considering 10, 20, and 50 different shopping profile types.

The experiments in Table~\ref{tab:noisy_data} demonstrate the impact of the attacker knowledge of the target data points by evaluating how adding varying degrees of to data features may impact on the success rate of membership inference attacks. Noise uniformly sampled from $[0,\sigma], \sigma \leq 1$ and added to features normalized within $[0,1]$. Given a level of uncertainty of $\mathbf{x}$ or inaccuracy in $D'$ on the part of the attacker, represented by a corresponding degree of noise $\sigma$, Table~\ref{tab:noisy_data} evaluates how effective the attacker remains in launching a membership inference attack to identify if $\mathbf{x} \in D$ or $\mathbf{x} \notin D$. The results reported in Table \ref{tab:noisy_data} are reported for four logistic regression models, each one trained on a different dataset with gradually increasing $\sigma$ values (degree of noise). 

We make two interesting observations from Table \ref{tab:noisy_data}. First, for all four datasets, the more accurate (the less noise) the attacker knowledge about $\mathbf{x}$ is, the higher the model vulnerability (attack success rate) to membership inference. This shows the accuracy of attacker knowledge about the targeted examples is an important factor in determining model vulnerability and attack success rate (in terms of attack accuracy). Second, in comparison to the noisy target data, adding noise to the shadow dataset results in a less severe drop in accuracy. Similar trends are however still observed with slightly higher attack success rates under smaller $\sigma$ values for all four datasets. This set of experiments demonstrates that attackers with different knowledge and different levels of resources may have different success rates in launching a membership inference attack. Thus, model vulnerability should be evaluated by taking into account potential or available attacker knowledge.

\subsection{Transferability of Membership Inference}

Inspired by the transferability of adversarial examples~\cite{szegedy2013intriguing},~\cite{papernot2016practical},~\cite{papernot2016transferability},~\cite{rozsa2016accuracy}, membership inference attacks are also shown to be transferable. That is, attack model $F_a$ trained on an attack dataset $D*$ containing the outputs from a set of shadow models is effective not only when shadow models and the target model are of the same type but also when the shadow model type varies. This property further opens the door to the black-box attackers who do not have any knowledge of the target model.
\begin{table}[!ht]
  \centering
  \resizebox{0.85\columnwidth}{!}{
  \begin{tabular}{|c||c|c|c|c|}
    \hline
    \textbf{Purchases-20}   & \multicolumn{4}{|c|}{Shadow Model Type} \\
    \hline
    Attack Model            & DT                & k-NN              & LR                & NB    \\
    \hline
    DT                      & \textbf{88.98}    & \textbf{87.49}    & 72.08             & 81.84 \\
    k-NN                    & \textbf{88.23}    & 72.57             & \textbf{84.75}    & 74.27 \\
    LR                      & \textbf{89.02}    & \textbf{88.11}    & \textbf{88.99}    & 83.57 \\
    NB                      & \textbf{88.96}    & 78.60             & \textbf{89.05}    & 66.34 \\
    \hline
  \end{tabular}\newline}
  \caption{\small{Accuracy (\%) of membership inference attack against a decision tree target model trained on the Purchases-20 dataset.}} 
  \label{tab:purch20_carttarget}
\end{table}

Table~\ref{tab:purch20_carttarget} demonstrates this property of membership inference attacks. It reports the membership inference attack accuracy for various attack configurations against a decision tree model trained on the Purchases-20 dataset. It shows the transferability of membership inference attacks for different combinations from four different model types uses as the attack model type $F_a$ (rows) and the shadow model type (columns). In this experiment, the target prediction model is a decision tree. We observe that while using decision tree as the shadow model results in the most consistent membership inference attack success compared to other combinations, multiple combinations with both $k$-NN and logistic regression (LR) shadow models achieve attack success within 5\% of gap compared to the most successful attack configuration using decision tree shadow models. Table~\ref{tab:purch20_carttarget} also shows that multiple types of models can be successful as the binary attack classifier $F_a$. This set of experiments also shows that the worst attack performances against the DT target prediction model are seen when the shadow models are trained using Na\"ive Bayes (NB), with the worst performance reported when NB is the model type of the attack model as well. These results indicate that (1) the same strategy used for selecting the shadow model type may not be optimal for the attack model and (2) shadow models of different types other than the target model type may still lead to successful membership inference attacks. We refer readers to~\cite{truex2019demystifying} for additional detail. 

The transferability study in Table~\ref{tab:purch20_carttarget} indicates an attacker does not always need to know the exact target model configuration to launch an effective membership inference attack as attack models can be transferable from one target model type to another. And although finding the \textit{most} effective attack strategy can be a challenging task for attackers, vulnerability to membership inference attack remains serious even with suboptimal attack configurations with almost all configurations reporting attack accuracy about $70\%$ and many above $85\%$.

\subsection{Training Data Skewness on Membership Attacks}

The fourth important dimension of membership inference vulnerability is the risk imbalance across different prediction classes when the training data is skewed. Even when the overall membership inference vulnerability appears limited with attack success close to the 50\% baseline (in or out random guess), there may be subgroups within the training data, which display significantly more vulnerability. 

\begin{figure}
  \centering
  \includegraphics[width=\columnwidth]{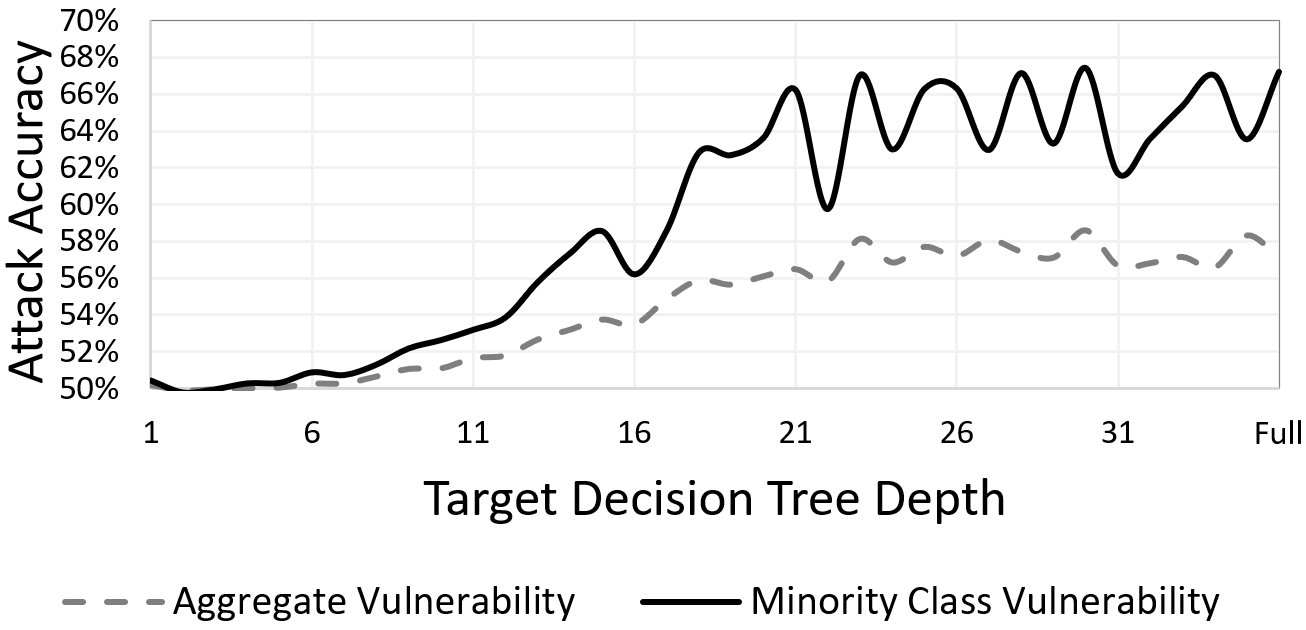}
  \caption{\small{Vulnerability of Adult dataset highlighting minority ($>$50K) class with decision trees.}}
  \label{fig:adult}
\end{figure}

For example, Figure~\ref{fig:adult} illustrates the impact of data skewness on membership inference vulnerability. In this set of experiments, we measure the membership inference attack accuracy for a decision tree target model trained on the publicly available Adult dataset~\cite{Dua:2017}. The adult dataset contains 48,842 instances, each with 14 different features and presents a binary classification problem wherein one wishes to identify if an individual's yearly salary is $> \$50$K or $\leq \$50$K. The class distribution, however, is skewed with less than $25\%$ of instances being labeled $>$\$50K. As overfitting is widely considered a key factor in membership inference vulnerability, we simulate overfitting by increasing the depth of the target decision tree model. Figure~\ref{fig:adult} shows that the impact of overfitting (increasing in X-axis) on both the aggregate membership inference vulnerability in terms of membership inference attack accuracy (accuracy over all classes) and the minority membership inference vulnerability (attack accuracy over the minority class). In this case aggregate vulnerability is the accuracy of the membership inference attack evaluated on an equal number of randomly selected examples seen by the target model (``in'') as as unseen (``out'') while the minority vulnerability reports the membership inference attack accuracy evaluated on only the subset of the previously selected examples whose class is $>$\$50K (the minority class).

We observe from Figure~\ref{fig:adult} that, the minority class has an increased risk under the membership inference attack as the model over-fits more severely. This follows the intuition that minority class members have fewer other instances amongst whom they can hide in the training set and thus are more easily exposed under membership inference attacks. This aligns well to some extent with the observation that smaller training dataset sizes can lead to a greater overall risk for membership inference~\cite{shokri2017membership}. We argue that it is important for both data owners for model training and the MLaaS providers to consider vulnerability not just for the entire training dataset, but also the level of risk for minority populations specifically when evaluating privacy compliance.

\section{Mitigation Strategies and Algorithms}

\subsection{Differential Privacy}
Differential privacy is a formal privacy framework with a theoretical foundation and rigorous mathematical guarantees when effectively employed~\cite{dwork2008differential}. A machine learning algorithm is defined to be differentially private if and only if the inclusion of a single instance in the training dataset will cause only statistically insignificant changes to the output of the algorithm. Theoretical limits are set on such output changes in the definition of differential privacy, which is given formally as follows:
\vspace{\baselineskip}
\begin{definition}[Differential Privacy~\cite{dwork2008differential}]
A randomized mechanism $K$ provides $(\epsilon, \delta)$-
differential privacy if for any two neighboring database $D_1$ and $D_2$
that differ in only a single entry, $\forall S \subseteq
Range(K)$,
\begin{equation}
\Pr(K(D_1) \in S) \le e^\epsilon \Pr(K(D_2) \in S) +
\delta
\end{equation}
\label{def:dp}
\end{definition}
If $\delta=0$, $K$ is said to satisfy $\epsilon$-differential privacy.
\vspace{\baselineskip}

In the remaining of the paper, we focus on $\epsilon$-differential privacy for presentation convenience. To achieve $\epsilon$-differential privacy (DP), noise defined by $\epsilon$ is added to the algorithm's output. This noise is proportional to the \textit{sensitivity} of the output. Sensitivity measures the maximum change of the output due to the inclusion of a single data instance.
\vspace{\baselineskip}
\begin{definition}[Sensitivity~\cite{dwork2008differential}]
For $f: \mathcal{D} \rightarrow
\mathbb{R}^k$, the sensitivity of $f$ is
\begin{equation}
\Delta = \max\limits_{D_1,D_2} ||f(D_1)-f(D_2) ||_2
\end{equation}
for all $D_1$, $D_2$ differing in at most one element.
\end{definition}

The noise mechanism which is used is therefore bounded by both the sensitivity of the function $f$, $S_f$, and the privacy parameter $\epsilon$. For example, consider the Gaussian mechanism defined as follows:
\begin{definition}[Gaussian Noise Mechanism]\label{def:gaussian}
$M(d) \triangleq f(D) + N(0, S^2_f\sigma^2)$
\end{definition}
where $N(0, S^2_f\sigma^2)$ is the normal distribution with mean $0$ and standard deviation $S_f\sigma$. A single application of the Gaussian mechanism in Definition~\ref{def:gaussian} to a function $f$ with sensitivity $S_f$ satisfies $(\epsilon, \delta)$-differential privacy if $\delta \geq \frac{5}{4}\textnormal{exp}(-(\sigma\epsilon)^2/2)$ and $\epsilon < 1$~\cite{dwork2014algorithmic}.

Additionally, there exist several nice properties of differential privacy for either multiple iterations of a differentially private function $f$ or the combination of multiple different functions $f_1, f_2, ...$ wherein each $f_i$ satisfies a corresponding $\epsilon_i$-differential privacy. These composition properties are important for machine learning processes which often involve multiple passes over the training dataset $D$. The formal composition properties of differential privacy include the following:

\begin{definition}[Composition properties~\cite{dwork2014algorithmic, mcsherry2009privacy}]\label{def:composition}
Let $f_1, f_2, \ldots, f_n$ be $n$ algorithms, such that for each $i \in [1, n]$, $f_i$ satisfies $\epsilon_i$-DP. Then, the following properties hold:
\begin{itemize}
    \item Sequential Composition: Releasing the outputs $f_1(D), f_2(D), \ldots, f_n(D)$ satisfies $(\sum_{i=1}^n \epsilon_i)$-DP.
    \item Parallel Composition: Executing each algorithm on a disjoint subset of $D$ satisfies $\textnormal{max}_i(\epsilon_i)$-DP.
    \item Immunity to Post-processing: Computing a function of the output of a differentially private algorithm does not deteriorate its privacy, e.g., publicly releasing the output of $f_i(D)$ or using it as an input to another algorithm does not violate $\epsilon_i$-DP.
\end{itemize}
\end{definition}

\vspace{\baselineskip}
Differential privacy can be employed to different types of machine learning models. Due to the space constraint, in the rest of the paper, we specifically focus on differentially private training of deep neural network (DNN) models.

\subsection{Mechanisms for Differentially Private Deep Learning}

DNNs are complex, sequentially stacked neural networks containing multiple layers of interconnected nodes. Each node represents the dataset in a unique way and each layer of networked nodes processes the input from previous layers using learned weights and a pre-defined activation function. The objective in training a DNN is to find the optimal weight values for each node in the multi-tier networks. This is accomplished by making multiple passes over the entire dataset with each pass constituting one \textit{epoch}. 
Within each epoch, the entire dataset is partitioned into many mini-batches of equal size and the algorithm processes these batches sequentially, each including only a subset of the data. When processing one batch, the data is fed forward through the network using the existing weight values. A pre-defined loss function is computed for the errors made by the neural network learner with respect to the current batch of data. An optimizer, such as stochastic gradient descent (SGD), is then used to propagate these errors backward through the network. The weights are then updated according to the errors and the learning rate set by the training algorithm. The higher the learning rate value, the larger the update made in response to the backward propagation of errors.

Differentially private deep learning can be implemented by adding a small amount of noise to the updates made to the network such that there is only a marginal difference between the following two scenarios: (1) when a particular individual is included within the training dataset and (2) when the individual is absent from the training dataset. The noise added to the updates is sampled from a Gaussian distribution with scale determined by an appropriate noise parameter $\sigma$ corresponding to a desired level of privacy and controlled sensitivity. That is, the privacy budget $\epsilon$, according to Definition~\ref{def:dp}, should constrain the value of $\sigma$ at each epoch. Let $n$ denote the number of epochs for the DNN training, a pre-defined hyper-parameter set at the training configuration as the termination condition. Let one epoch satisfy $\epsilon_i$-differential privacy given $\sigma_i$ from the Gaussian mechanism in Definition~\ref{def:gaussian}. Then, a traditional accounting using the composition properties of differential privacy (Definition~\ref{def:composition}) would dictate that $n$ epochs would result in an overall privacy guarantee of $\epsilon$ if $\epsilon_i = \epsilon / n$, and each epoch employed the Gaussian mechanism with the same value $\sigma_i$. We refer to this approach the fixed noise perturbation method~\cite{abadi2016deep}. An alternative approach proposed by Yu et. al. in~\cite{yu2019differentially} advocates a variable noise perturbation approach, which uses a decaying function to manage the total privacy budget $\epsilon$ and define variable noise scale $\sigma$ based on different settings of $\epsilon_i$ for each different epoch $i$ ($1\leq i \leq n$), aiming to add variable amount of noise to the $i^{th}$ epoch in a decreasing manner as the training progresses in epochs. Thus, we have $\epsilon_{i} \neq \epsilon_{j}$ for $i\neq j$, $1\leq i, j\leq n$ and $\sigma_i$ for a given epoch $i$ is bounded by its allocated privacy budget $\epsilon_i$. The same overall privacy guarantee is met when $\sum_{i=1}^n \epsilon_i, = \epsilon$ for the differentially private DNN training of $n$ epochs.

\subsection{Differentially Private Deep Learning with Fixed $\sigma$}

The first differentially private approach to training deep neural networks is proposed by Abadi et al.~\cite{abadi2016deep} and implemented on the tensorflow deep learning framework~\cite{tensorflow2015-whitepaper}. A summary of their approach is given in Algorithm~\ref{algo:abadi}. To apply differential privacy, the sensitivity of each epoch is bounded by a clipping value $C$, specifying that an instance may impact weight updates by at most the value $C$. To achieve differential privacy, weight updates at the end of each batch include noise injection according to the sensitivity defined by $C$ and the scale of noise $\sigma$. The choice of $\sigma$ is directly related to the overall privacy guarantee. 

\begin{algorithm}
\caption{Differentially Private Deep Learning: Fixed $\sigma$}\label{algo:abadi}
\begin{algorithmic}
\State \textbf{Input}: Dataset $D$ containing training instances $x_1, ..., x_N$, loss function $L(\theta) = 1/N \sum_i L(\theta, x_i)$, learning rate $\eta_t$, noise scale $\sigma$, batch size $L$, norm bound $C$, number of epochs $E$
\State \textbf{Initialize} $\theta_0$ randomly
\State \textbf{Set} $T = E * N/L$
\State \textbf{for} $t \in [T]$ \textbf{do}
\State \hskip1em Set sampling probability $q = L/N$
\State \hskip1em Take a random sample $L_t$ from $D$
\State \hskip1em \textbf{Compute gradient}
\State \hskip1em For each $x_i \in L_t$, compute $\mathbf{g}_t(x_i) \leftarrow \nabla_{\theta_t} L(\theta_t, x_i)$
\State \hskip1em \textbf{Clip gradient}
\State \hskip1em $\mathbf{\bar{g}}_t(x_i) \leftarrow \mathbf{g}_t(x_i) / \max (1, \frac{\|\mathbf{g}_t(x_i)\|_2}{C})$
\State \hskip1em \textbf{Add Noise}
\State \hskip1em $\mathbf{\Tilde{g}}_t \leftarrow 1/L (\sum_i \mathbf{\bar{g}}_t(x_i) + N(0, \sigma^2C^2\text{\textbf{I}}))$
\State \textbf{Descent}
\State $\theta_{t+1} \leftarrow \theta_t - \eta_t\mathbf{\Tilde{g}}_t$
\State \textbf{Output} $\theta_T$ and compute the overall privacy cost $(\epsilon, \delta)$ using a privacy accounting method.
\end{algorithmic}
\end{algorithm}

Let $\sigma = \frac{\sqrt{2\log(1/\delta)}}{\epsilon}$, then according to differential privacy theory~\cite{dwork2014algorithmic}, each step (processing of a batch) is $(\epsilon, \delta)$-differentially private. If $L_t$ is randomly sampled from $D$ then additional properties of random sampling \cite{kasiviswanathan2011can, beimel2010bounds} may be applied. Each step then becomes $(O(q\epsilon), q\delta)$-differentially private. The moments accountant privacy accounting method is also introduced in~\cite{abadi2016deep} to prove that Algorithm~\ref{algo:abadi} is $(O(q\epsilon\sqrt{T}), \delta)$-differentially private given appropriate parameter settings.

We refer to Algorithm~\ref{algo:abadi} as the fixed noise perturbation approach as each epoch is treated equally by introducing the same noise scale to every parameter update.

\subsection{Differentially Private Deep Learning with Variable $\sigma$}

The variable noise perturbation approach to differentially private deep learning is proposed by Yu et al.~\cite{yu2019differentially}. It extends the fixed noise scale of $\sigma$ over the total $n$ epochs in~\cite{abadi2016deep} by introducing two new capabilities. First, Yu et al. in \cite{yu2019differentially} pointed out a limitation of the approach outlined in Algorithm \ref{algo:abadi}. Namely, Algorithm \ref{algo:abadi} specifically calls for random sampling, wherein each batch is selected randomly with replacement. However, the most popular implementation for partitioning a dataset into mini-batches in many deep learning frameworks is random shuffling, wherein the dataset is shuffled and then partitioned into evenly sized batches. In order to develop a differentially private DNN model under random shuffling, Yu et.al~\cite{yu2019differentially} extends Algorithm \ref{algo:abadi} of~\cite{abadi2016deep} by introducing a new privacy accounting method.

Additionally, Yu et al.~\cite{yu2019differentially} analyze the problem of using fixed noise scale (fixed $\sigma$ values), and propose employing different noise scales to the weight updates at different stages of the training process. Specifically, Yu et al.~\cite{yu2019differentially} propose a set of methods for privacy budget allocation, which improve model accuracy by progressively reducing the noise scale as the training progresses. The variable $\sigma$ noise scale approach is inspired by two observations. First, as the training progresses, the model begins to converge causing the noise being introduced to the updates to potentially become more impactful. This slows down the rate of model convergence and causes later epochs to no longer increase model accuracy compared to non-private scenarios. Second, the research on improving training accuracy and convergence rate of DNN training has led to a new generation of learning rate functions that replace the constant learning rate baseline by decaying learning rate functions and cyclic learning rates~\cite{wu2019demystifying, chollet2017xception}. Yu et al.~\cite{yu2019differentially} employed a similar set of decay functions to add noise at a decreased scale. That is, the noise defined by $\sigma_i$ at the $i^{th}$ epoch as the training progresses is less than $\sigma_j$ at the $j^{th}$ epoch given $1 \leq j < i \leq n$. The performance of four different types of decay functions to introduce variable noise scale by partitioning $\sigma$ over $n$ epochs were evaluated.

\subsection{Important Implementation Factors}

\subsubsection{Choosing Epsilon}

In differentially private algorithms, the $\epsilon$ value dictates the amount of noise which must be introduced into the DNN model training and therefore the theoretical privacy bound. Choosing the correct $\epsilon$ value requires a careful balance between tolerable privacy leakage given the practical setting as well as the tolerable utility loss. 

For example, Naldi and D'Acquisto propose a method in~\cite{naldi2015differential} for finding the optimal $\epsilon$ value to meet a certain accuracy for Laplacian mechanisms. Lee and Clifton alternatively employ the approach in~\cite{lee2011much} to analyze a particular adversarial model. Hsu et. al~\cite{hsu2014differential}, on the other hand, takes an individual's perspective on data privacy by opt-in incentivization. Kohli and Laskowski~\cite{kohli2018epsilon} also promote choosing an $\epsilon$ based on individual privacy preferences.

Despite these existing approaches, determining the ``right'' $\epsilon$ value remains a complex problem and is likely to be highly dependent on the privacy policy of the organization that owns the model and the dataset, the vulnerability of the model, the sensitivity of the data, and the tolerance to utility loss in the given setting. Additionally, there might be scenarios, such as the healthcare setting, in which even small degrees of utility loss are intolerable and are combined with stringent privacy constraints given highly sensitive data. In these cases, it may be hard or even impossible to find a good $\epsilon$ value for existing differentially private DNN training techniques. 

\subsubsection{The Role of Transfer learning}

Another key consideration is the use of transfer learning for dealing with more complex datasets~\cite{abadi2016deep,yu2019differentially}. For example, model parameters may be initialized by training on a non-private, similar dataset. The private dataset is then only used to further hone a subset of the model parameters. This helps to reduce the number of parameters affected by the noise addition required by exercising differential privacy. The use of transfer learning however relies on a strong assumption that such a non-private, similar dataset exists and and is available. In many cases, this assumption may be unrealistic.

\subsubsection{Parameter Optimization}

In additional to the privacy budget parameters $(\epsilon, \delta)$, differentially private deep learning introduces influential privacy parameters, such as the clipping value $C$ that bounds the sensitivity for each epoch and the noise scale approach (including potential decay parameters) for $\sigma$. The settings of these privacy parameters may impact both the training convergence rate, and thus the training time, and the training and testing accuracy. However, tuning these privacy parameters becomes much more complex as we need to take into account the many learning hyper-parameters already present in DNN training, such as the number of epochs, the batch size, the learning rate policy, and the optimization algorithm. These hyper-parameters need to be carefully configured for high performance training of deep neural networks in a non-private setting. For deep learning with differential privacy, one needs to configure the privacy parameters by considering the side effect on other hyper-parameters and re-configure the previously tuned learning hyper-parameters to account for the privacy approach.

For example, for a fixed total privacy budget values $(\epsilon, \delta)$, too small of a number of epochs may result in reduced accuracy due to insufficient time to learn. A higher number of epochs however will result in a higher $\sigma$ value required each epoch (recall Algorithm~\ref{algo:abadi}). Therefore, a carefully tuned hyper-parameter for a non-private setting, such as the optimal number of epochs, may no longer be effective when differentially private deep learning is enabled. 

A number of challenging questions remain open problems in differentially private DNN training, such as at what point do more epochs lead to accuracy loss under a particular privacy setting? What is the right noise decaying function for effective deep learning with differential privacy? Can we learn privately with high accuracy given complex datasets? The balance of the many parameters in a differentially private deep learning system presents new challenges to practitioners.

With these questions in mind, we develop MPLens, a membership privacy analysis system, which facilitates the evaluation of model vulnerability against membership inference attacks. Through MPLens, we investigate the effectiveness of differential privacy as a mitigation technique against membership inference attacks, including the trade-offs of implementing such a mitigation strategy for preventing membership inference and the impact of differential privacy on different classes for the DNN models trained using skewed training datasets. We also highlight how the vulnerability of pre-trained models under the membership inference attack is not uniform when the training data itself is skewed with minority populations. We show how this vulnerability variation may cause increased risks in federated learning systems.

\begin{figure*}
  \centering
  \includegraphics[width=0.9\textwidth]{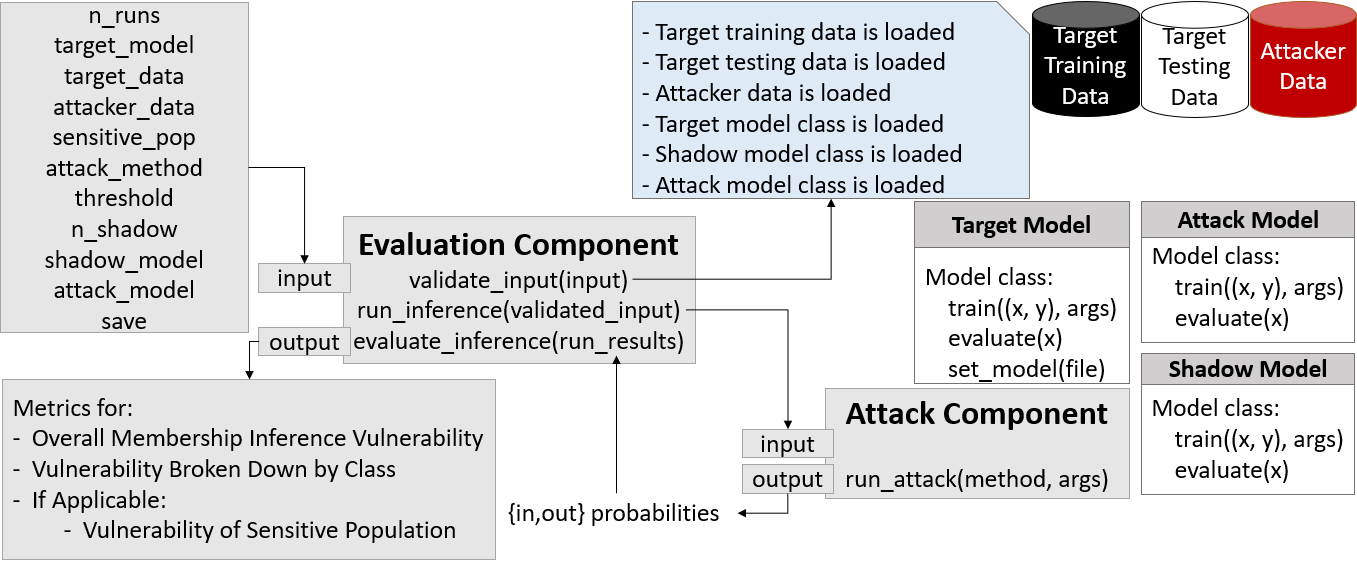}
  \caption{\small{Overview of System to Evaluate Membership Inference Vulnerability}}
  \label{fig:system}
  \vspace{-\baselineskip}
\end{figure*}

\section{MPLens: System Overview}

MPLens is designed as a privacy analysis and privacy compliance evaluation system for both data scientists and MLaaS providers to understand the membership inference vulnerability risk involved in their model training and model prediction process. Figure \ref{fig:system} provides an overview of the system architecture. The system allows providers to specify a set of factors that are critical to privacy analysis. Example factors include the data used to train their model, what data might be held by the attacker, what attack technique might be used, the degree of data skewness in the training set, whether the prediction model is constructed using the differentially private model training, what configurations are used for the set of differential privacy parameters, and so forth. Given the model input, the MPLens evaluation system reports the overall statistics on the vulnerability of the model, the per-class vulnerability, as well as the vulnerability of any sensitive populations such as specific minority groups. Example statistics include attack accuracy, precision, recall, and f-1 score. We also include attacker confidence for true positives, false positives, true negatives, and false negatives, the average distance from both the false positives and the false negatives to the training data, and the time required to execute the attack. 

\subsection{Target Model Training}

Over-fitting is the first factor that MPLens measures for conducting membership vulnerability analysis. MPLens specifically highlights the overfitting analysis by reporting the Accuracy Difference between the target model training accuracy and testing accuracy. This enables MLaaS providers and domain-specific data scientists to understand whether their vulnerability might be linked to overfitting. As previously indicated, while overfitting is strongly correlated with membership inference vulnerability, it is not the only source of vulnerability. Thus, when MPLens indicates undesirable vulnerability an absence of significant overfitting, analysis may be triggered to investigate if vulnerability is linked to other model or data characteristics as those discussed in Section III. 

\subsection{Attacker Knowledge}

Our MPLens system is by design customizable to understand multiple attack scenarios. For instance, users may specify the shadow data, which is used by the attacker. This allows the user to consider a scenario in which the attacker has access to some subset of the target model's training data, one where the attacker has access to some data that are drawn from the same distribution as the training data of the target model, or one where the attacker has noisy and inaccurate data, such as that evaluated in Table~\ref{tab:noisy_data} and possibly generated through black-box probing~\cite{shokri2017membership, truex2019demystifying}. 

The MPLens system is additionally customizable with respect to the attack method, including the shadow model based attack techniques~\cite{shokri2017membership}, and the threshold-based attack techniques~\cite{yeom2018privacy}. Furthermore, when using a threshold-based attack, our MPLens system can either accept pre-determined values representing attacker knowledge of the target model error or it can also determine good threshold values through the shadow model training.

These customizations allow MLaaS providers and users of MPLens to specify the types of attackers they wish to analyze, analyze their model vulnerability against such attackers, and evaluate the privacy compliance of their model training and model prediction services.

\subsection{Transferability}

Another aspect of privacy analysis is related to the specific model training methods used to generate membership inference attacks and whether different methods result in significant variations in membership inference vulnerability. Consider the attack method from \cite{shokri2017membership}, the user can specify not only the attacker's data but also the shadow model training algorithm and the membership inference binary classifier training algorithm. Each element is customizable as a system parameter when configuring MPLens for specific privacy risk evaluation. The MPLens system makes no assumption on how the attacker develops the shadow dataset, what knowledge is included in the data, whether the attacker has knowledge of the target model algorithm, or what attack technique is used. This flexibility allows MPLens to support evaluation across various transferable attack configurations.


\begin{figure*}[ht]
  \centering
  \begin{minipage}{0.32\textwidth}
    \centering
    \includegraphics[width=0.8\linewidth]{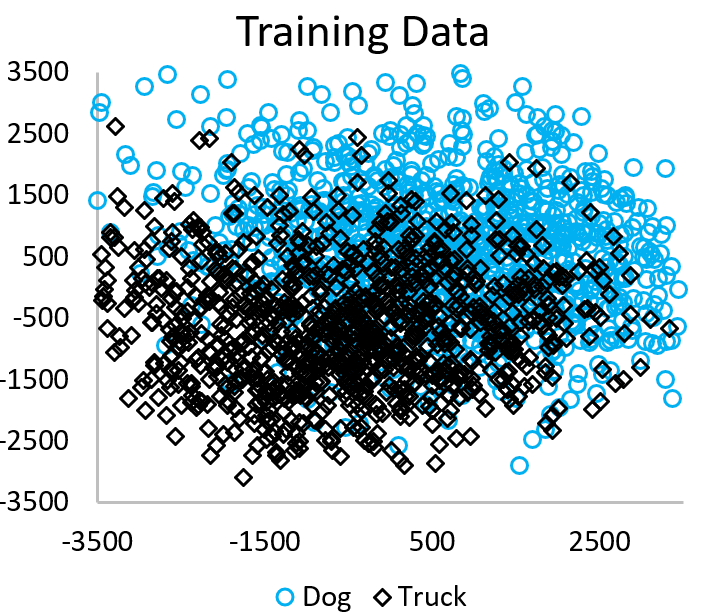}
    \caption{\small{Training Data}}
    \label{fig:pca_in}
  \end{minipage}
  \begin{minipage}{0.32\textwidth}
    \centering
    \includegraphics[width=0.8\linewidth]{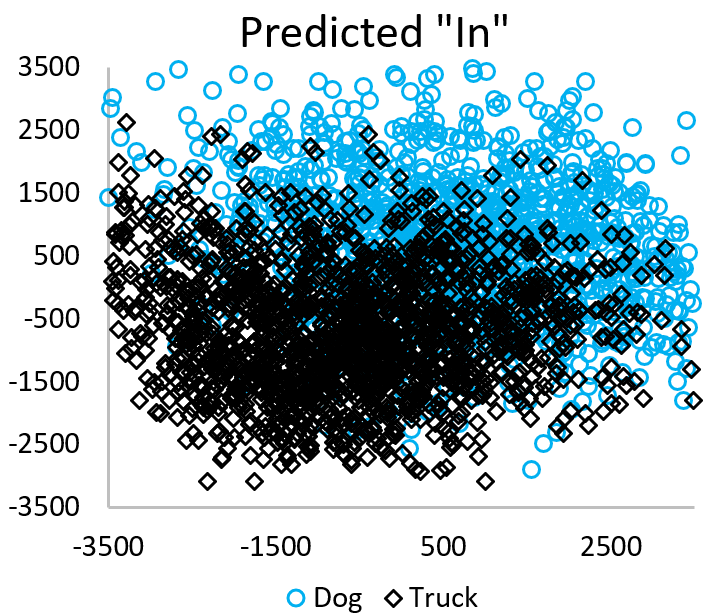}
    \caption{\small{Predicted In}}
    \label{fig:pca_pred_in}
  \end{minipage}
  \begin{minipage}{0.32\textwidth}
    \centering
    \includegraphics[width=0.8\linewidth]{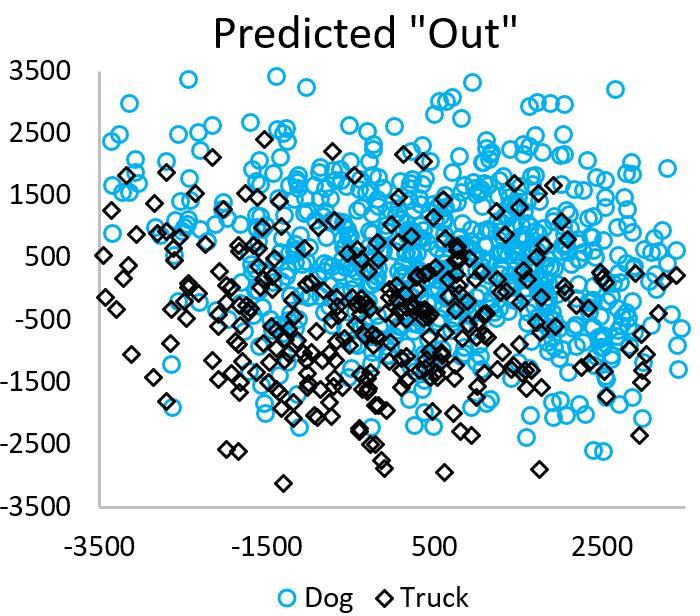}
    \caption{\small{Predicted Out}}
    \label{fig:pca_pred_out}
  \end{minipage}
  \captionsetup{labelformat=empty}
  \caption{\small{Comparison of 2-D PCA for CIFAR-10 dog and truck classes. Plots show (1) training data, (2) data predicted as training data, and (3) data predicted as non-training data by the membership inference attack.}}
  \vspace{-\baselineskip}
\end{figure*} \addtocounter{figure}{-1}

\section{Experimental Results and Analysis}

\subsection{Datasets}

All experiments reported in this section were conducted using the following four datasets.



\paragraph{CIFAR-10}

The CIFAR-10 dataset contains 60,000 color images~\cite{krizhevsky2009learning} and is publicly available. Each image is formatted to be 32 x 32. The CIFAR-10 dataset contains 10 classes with 6,000 images each: airplane, automobile, bird, cat, deer, dog, frog, horse, ship and truck. The problem is therefore a 10-class classification problem where the task is to identify which class is depicted in a given image.

\paragraph{CIFAR-100}
Also publicly available, CIFAR-100 similarly contains 60,000 color images~\cite{krizhevsky2009learning} formatted to 32 x 32. 100 classes are represented ranging from various animals, pieces of household furniture, or types of vehicles. Each class has 600 available images. The problem is therefore a 100-class image classification problem.

\paragraph{MNIST}

MNIST is a publicly available dataset containing 70,000 images of handwritten digits~\cite{lecun2010mnist}. Each image is formatted to be 32 x 32 and processed such that the digit is at the center of the image. The MNIST dataset constitutes a 10-class classification problem where the task is to identify which digit between $0$ and $9$, inclusive, is contained within a given image.

\paragraph{Labeled Faces in the Wild}

The Labeled Faces in the Wild (LFW) database contains face photographs for unconstrained face recognition with more than 13,000 images of faces collected from the web. Each face has been labeled with the name of the person pictured. 1,680 of the people pictured have two or more distinct photos in the data set. Each person is then labeled with a gender and race (including mixed races). Data is then selected for the top 22 classes which were represented with a sufficient number of data points.

\subsection{Membership Inference Risk: Adversarial Examples}

Given a target model and its training data, the membership inference attack, using black-box access to the model prediction API, can be used to create a representative dataset. This representative dataset can be leveraged to generate substitute models for the given target model. One can then use such substitute models to generate adversarial examples using different adversarial attack methods~\cite{wei2019cross}. 

Figures \ref{fig:pca_in}-\ref{fig:pca_pred_out} provide visualization plots for the comparison of 2-D PCA given the images of the dog and truck classes in CIFAR-10. The plots are divided relative to the membership inference attack prediction output including the true target model training data (Figure~\ref{fig:pca_in}), the data predicted by the membership inference attack as training data (Figure~\ref{fig:pca_pred_in}), and the data predicted as non-training data by the membership inference attack (Figure~\ref{fig:pca_pred_out}).

These plots illustrate the accuracy of the distribution of instances predicted as in the target model's training dataset through membership inference. They clearly demonstrate how even with the inclusion of false positives, an attacker can create a good representation of the training data distribution, particularly compared to those instances not predicted to be in the target training data. An attacker can therefore easily train a substitute model on this representative dataset which then enables the generation of adversarial examples by attacking this substitute model. Examples developed to successfully attack a substitute model trained on the instances in Figure~\ref{fig:pca_in} are likely to also be successful against a model trained on the instances in Figure~\ref{fig:pca_in}.

\begin{figure}
  \centering
  \includegraphics[width=\columnwidth]{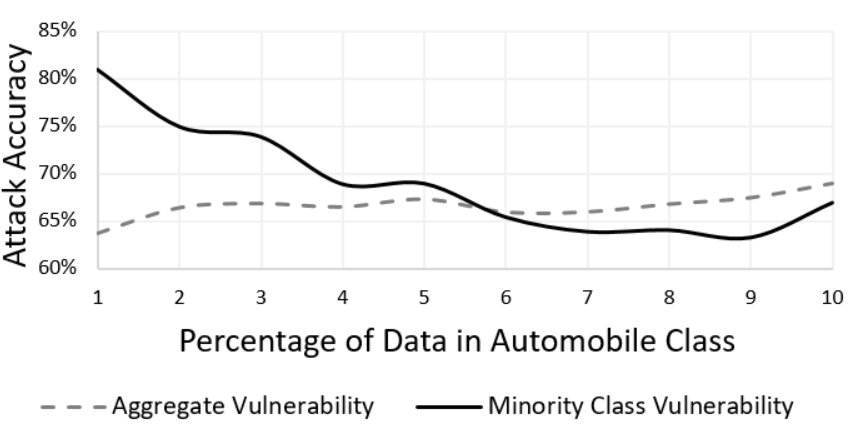}
  \caption{\small{Minority Class Percentage vs Membership Inference Vulnerability. Experiments are conducted with the CIFAR-10 dataset with the automobile as the minority class.}}
  \label{fig:cifar10_distrib}
\end{figure}

\subsection{Membership Inference Risk: Data Skewness}

To date, the membership inference attack has been studied either using training datasets with uniform class distribution or without specific consideration of the impact of any data skewness. However, as we demonstrated earlier, the risk of membership inference vulnerability can vary when class representation is skewed. Minority classes can display increased risk to membership inference attack as models struggle to more effectively generalize past the training data when fewer instances are given. In this section, we focus on studying the impact of data skewness on vulnerability to membership inference attacks. 

In Figure~\ref{fig:cifar10_distrib} we investigate the impact of data skewness on membership inference vulnerability by controlling the representation of a single class. We reduce the automobile images from the CIFAR-10 dataset to only 1\% of the data and then increase the representation until the dataset is again balanced with automobiles representing 10\% of the training images. We then plot the aggregate membership inference vulnerability which is the overall membership inference attack accuracy evaluated across all classes as well as the vulnerability of just the automobile class.

Figure~\ref{fig:cifar10_distrib} demonstrates that in cases where the automobile class constitutes 5\% or less of the total training dataset, i.e. the automobile class has fewer than half as many instances as each of the other classes, this skewness will result in the automobile class displaying more severe vulnerability to membership inference attack.

Interestingly, the automobile class displays lower membership inference vulnerability than the average vulnerability reported in the CIFAR-10 dataset when the dataset is balanced. However, when the automobile class becomes a minority class with fewer than half the instances of each of the other classes, the vulnerability shifts to be greater than that reported by the overall model. This gap becomes greater as continued decreased representation results in continued increased vulnerability for the automobile class.

\begin{table}[!ht]
  \centering
  \resizebox{0.85\columnwidth}{!}{
  \begin{tabular}{|c|c|}
    \hline
    Target Population & Attack Accuracy (\%) \\
    \hline
    \hline
    Aggregate & 70.14 \\ 
    \hline
    Male Images & 68.18 \\ 
    \hline
    Female Images & \textbf{76.85} \\ 
    \hline
    White Race Images & 62.77 \\ 
    \hline
    Racial Minority Images & \textbf{89.90} \\ 
    \hline
  \end{tabular}}
  \caption{\small{Vulnerability to membership inference attacks for different subsets of the LFW dataset.}}
  \label{tab:lfwa_minority}
\end{table}

Table~\ref{tab:lfwa_minority} additionally shows the vulnerability of a DNN target model trained on the LFW dataset to membership inference attacks. We analyze this vulnerability by breaking down the aggregated vulnerability across the top 22 classes into four different (non-disjoint) subsets of the LFW dataset: Male, Female, White Race, and Racial Minority. We observe that the training examples of racial minorities experience the highest attack success rate ($89.90\%$) and are thus highly vulnerable to membership inference attacks compared to images of white individuals. Similarly, female images, which represent less than $25\%$ of the training data, demonstrate higher average vulnerability ($76.85\%$) compared with images of males ($68.18\%$). 

To provide deeper insight and more intuitive illustration for the increased vulnerability of minority groups under membership inference attacks, Table~\ref{tab:lfw_examples} provides 7 individual examples of images in the LFW dataset targeted by the membership inference attack. That is, given a query with each example image, the target model predicts the individual's race and gender (22 separate classes) which the attack model, a binary classifier, uses to predict if that image was ``in'' or ``out'' of the target model's training dataset. The last row reports the ground truth as to whether or not the image was in the target model training set. 

\begin{table*}[!ht]
  \centering
  \begin{tabular}{|c|c|c|c|c|c|c|c|}
  \hline
     \makecell{\textcolor{ForestGreen}{\ding{51}} = correct\\ prediction\\ 
     \textcolor{red}{\ding{55}} = wrong\\ prediction} & \raisebox{-.7\totalheight}{\includegraphics[width=0.095\textwidth]{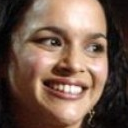}} &\raisebox{-.7\totalheight}{\includegraphics[width=0.095\textwidth]{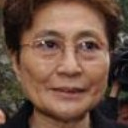}} & \raisebox{-.7\totalheight}{\includegraphics[width=0.095\textwidth]{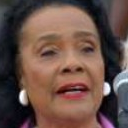}} & \raisebox{-.7\totalheight}{\includegraphics[width=0.095\textwidth]{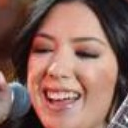}} & \raisebox{-.7\totalheight}{\includegraphics[width=0.095\textwidth]{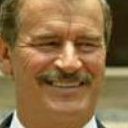}} & \raisebox{-.7\totalheight}{\includegraphics[width=0.095\textwidth]{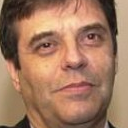}} & \raisebox{-.7\totalheight}{\includegraphics[width=0.095\textwidth]{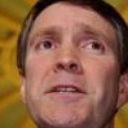}} \\
     & & & & & & & \\
     \hline
     \hline
     Target & \multirow{2}{*}{\textcolor{ForestGreen}{\ding{51}} 99.99} & \multirow{2}{*}{\textcolor{ForestGreen}{\ding{51}} 65.81} & \multirow{2}{*}{\textcolor{ForestGreen}{\ding{51}} 72.56} & \multirow{2}{*}{\textcolor{red}{\ding{55}} 62.30} & \multirow{2}{*}{\textcolor{ForestGreen}{\ding{51}} 99.99} & \multirow{2}{*}{\textcolor{red}{\ding{55}} 99.63} & \multirow{2}{*}{\textcolor{ForestGreen}{\ding{51}} 98.38} \\
     Confidence (\%) & & & & & & & \\
     \hline
     Attacker & \multirow{2}{*}{\textcolor{ForestGreen}{\ding{51}} 86.10} & \multirow{2}{*}{\textcolor{ForestGreen}{\ding{51}} 50.49} & \multirow{2}{*}{\textcolor{red}{\ding{55}} 61.85} & \multirow{2}{*}{\textcolor{ForestGreen}{\ding{51}} 72.06} & \multirow{2}{*}{\textcolor{ForestGreen}{\ding{51}} 56.40} & \multirow{2}{*}{\textcolor{ForestGreen}{\ding{51}} 99.88} & \multirow{2}{*}{\textcolor{red}{\ding{55}} 53.29} \\
     Confidence (\%) & & & & & & & \\
     \hline
     In Training & \multirow{2}{*}{\textcolor{ForestGreen}{in}} & \multirow{2}{*}{\textcolor{red}{out}} & \multirow{2}{*}{\textcolor{red}{out}} & \multirow{2}{*}{\textcolor{red}{out}} & \multirow{2}{*}{\textcolor{ForestGreen}{in}} & \multirow{2}{*}{\textcolor{red}{out}} & \multirow{2}{*}{\textcolor{red}{out}} \\
     Data? & & & & & & & \\
     \hline
  \end{tabular}
  \caption{\small{Examples of target and attack model performance and confidence on various images from the LFW dataset.}}
  \label{tab:lfw_examples}
  \vspace{-1.5\baselineskip}
\end{table*}

Through Table~\ref{tab:lfw_examples}, we highlight how minority populations are more likely to be identified by an attacker with a higher degree of confidence. We next discuss each example from left to right to articulate the impact of data skewness on model vulnerability to the membership inference attack.

For the 1st image, the target model is highly confident with its prediction and its prediction is indeed correct. Using the membership inference attack model, the attacker predicts that the model must have seen this example with high confidence and it succeeds in the membership inference attack.

For the 2nd image, the target model is less confident with its prediction, although the prediction outcome is correct. The attacker succeeds in the membership inference attack because it correctly predicts that this example is not in the training set, though the attacker’s confidence on this membership inference is much less certain (close to $50\%$) compared to that of the attack to the 1st image. We conjecture that the relatively low prediction confidence by the target model may likely contribute to the fact that the attacker is unable to obtain a high confidence for his membership inference attack.

The 3rd image is predicted by the target model correctly with a confidence of 72.56\%, which is about 11.5\% more confidence than that for the 2nd image. However, the attacker wrongly predicts that the example is in the training set when the ground truth shows that this example is not in the training set. Assuming the same logic as with the 1st image, i.e., the confidence and accuracy of target model prediction may indicate that the image was in the training dataset, could have caused the attacker to be misled.

For the 4th image, the target model has an incorrect prediction with the confidence of 62.30\%. The attacker correctly predicts that this example is not in the training set. It is clear that a somewhat confident and yet incorrect prediction by the target model is likely to result in high attacker confidence that this minority individual has not been seen during the training.

These four images highlight the compounding downfall for minority populations. Models are more likely to overfit these populations. This leads to poor test accuracy for these populations \textit{and} makes them more vulnerable to attack. As the 3rd example shows, the way to fool attackers is to have an accurate target model that can show reasonable confidence when classifying minority test images.

We next compare the results from the previous four images which represented minority classes with results from three images representing the majority class (white male images). For the 5th image in Table~\ref{tab:lfw_examples}, the target model predicts correctly with high confidence and the attacker is able to correctly predict that the image was in the training data. This result can be interpreted through comparison with the attacker performance for the 1st image. For the 5th query image which is from the majority group, the attacker predicts correctly that the image has been seen in training with barely over 50\% in confidence, showing relatively high uncertainty compared with the 1st image from a minority class. This indicates that model accuracy and confidence are weaker indicators with respect to membership inference vulnerability for the majority class.

For the 6th image, the target model produces an incorrect prediction with high confidence. The attacker is very confident that the query image is not in the training dataset, which is indeed the truth. This demonstrates rare a potential vulnerability for the majority class: When the target model has high confidence in an inaccurate prediction, an attack model is able to confidently succeed in the membership inference attack. Through this example and the above analysis, we see that the majority classes have two advantages compared to the minority classes: (1) it is less common for the target model to demonstrate this vulnerability of misclassification with high confidence; and (2) for majority classes, the accuracy and privacy are aligned rather than as competing objectives.

For the 7th image, the target model makes a correct prediction with high confidence. The attacker makes the incorrect prediction that the example was in the training dataset of the target model. But the truth is that the example is not in the training set. This membership attack failed and is the flip side of the 5th image, with both reporting low attack confidence. Again by comparing with the 1st image of minority, it shows how model confidence and accuracy may lead to membership inference vulnerability for the minority classes in a way that is not true for the majority classes.

\subsection{Mitigation with Differentially Private Training}

The second core component of our experimental analysis is to use MPLens to investigate the effectiveness of differential privacy employed to deep learning models as a countermeasure for membership inference mitigation. 

To define utility loss we follow~\cite{jayaraman2019evaluating} and consider $loss = 1 - \frac{dp-acc}{acc}$ where $acc$ represents accuracy in a non-private setting and $dp-acc$ is the accuracy when differential privacy is employed for the same model and data.

\subsubsection{{\bf Model and Problem Complexity}}\hfill\\
\vspace{-\baselineskip}

In Table~\ref{tab:dputil_vs_mivulnerable} compares three different datasets (first column) and learning tasks (third column) using different model structures (second column). We measure utility loss incurred when differential privacy is introduced, including both the training loss and test loss, as well as the vulnerability when non-private training is used. 
For MNIST, we follow the architecture outlined in~\cite{abadi2016deep} by first applying PCA with 60 components to reduce data dimensionality before feeding the data into simple neural network with one hidden ReLU layer containing 1,000 units. 

For CIFAR-10 and CIFAR-100 we test two separate training approaches and architectures. 
First, we implement the architecture used in~\cite{jayaraman2019evaluating} for both CIFAR-10 and CIFAR-100. This includes no transfer learning but does include a PCA transformation prior to training to reduce dimensionality to 50 features. The network used in this case consists of two fully connected layers with 256 units each and a softmax layer for the output. 

We also follow the approach from~\cite{abadi2016deep} which uses two ReLU convolutional layers with 64 channels, $5\times5$ convolutions, and a stride of 1. Each convolutional layer is then followed by a $2\times2$ max pool. There are two fully connected layers with 384 units before the final softmax layer. Prior to training the network, transfer learning (TL) is used to initialize model weights. For CIFAR-10 transfer learning is done with the CIFAR-100 dataset while the reverse is done for the CIFAR-100 dataset. The convolutional layers are fixed so that only the fully connected and softmax layers are updated during training. Finally, the data is cropped to $24\times24$ to further reduce dimensionality. 

For the non-private setting, we fix learning to $100$ epochs and use the \textsc{RMSprop} optimizer with learning rate set to $0.001$, decay to $1e-6$, and the batch size set at $32$. For the differential privacy setting we use some of the default parameter values from the tensorflow privacy repository for MNIST example~\cite{tensorflowprivacy}: setting the learning rate to $0.15$, batch size to $256$, microbatches to $256$, and the clipping value to $1.0$. We then set the noise multiplier to the correct scale according to the privacy budget $(\epsilon, \delta)$ and $100$ epochs of learning. In Table~\ref{tab:dputil_vs_mivulnerable}, $(\epsilon, \delta) = (10, 10^{-5})$. We employ the R\'enyi accounting approach~\cite{geumlek2017renyi} and leverage the tensorflow \textsc{DPGradientDescentGaussianOptimizer}.

\begin{figure*}[!ht]
  \centering
  \begin{minipage}{0.49\textwidth}
    \centering
    \includegraphics[width=0.8\linewidth]{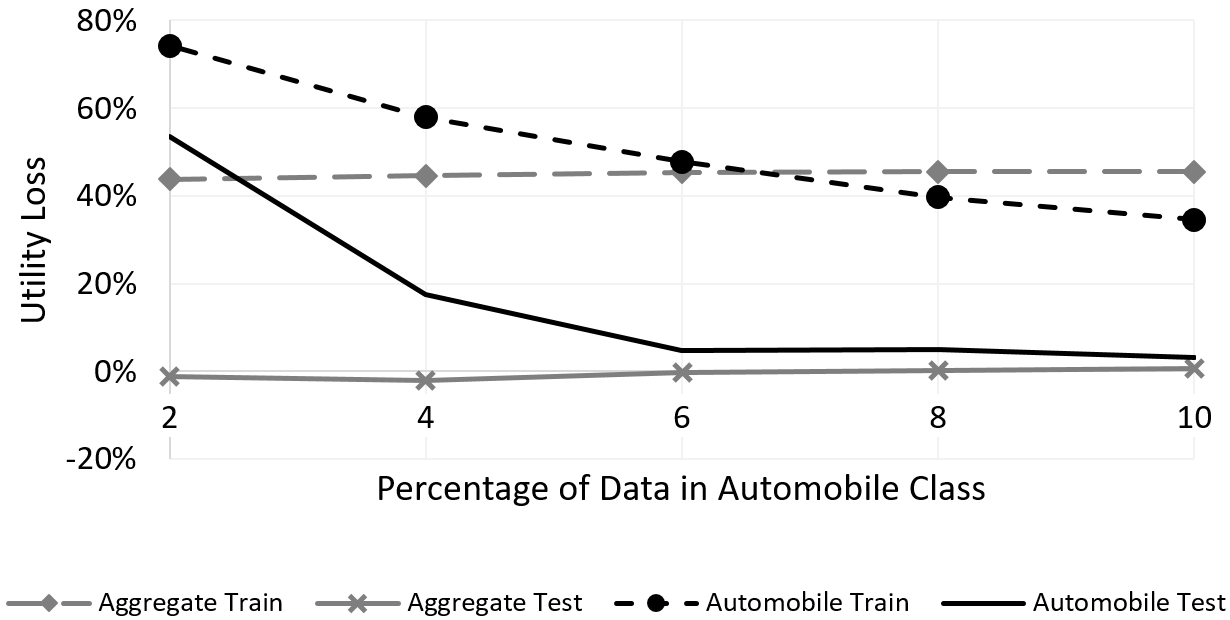}
  \end{minipage}%
  \begin{minipage}{0.49\textwidth}
    \centering
    \includegraphics[width=0.8\linewidth]{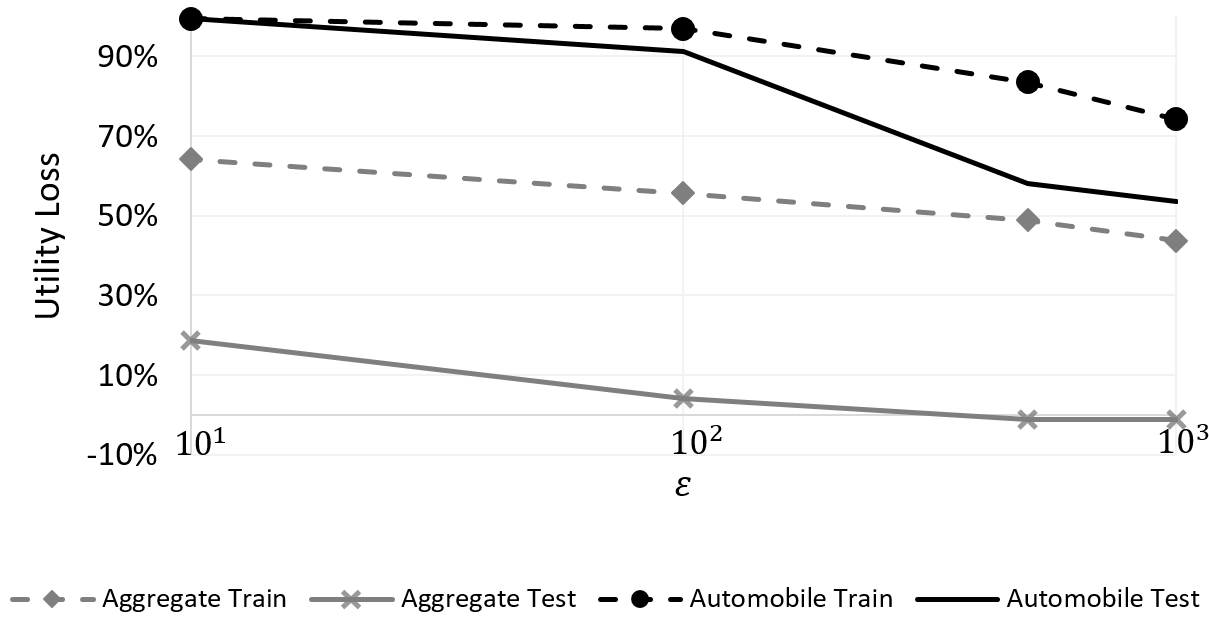}
  \end{minipage}%
  \caption{\small{Impact of differentially private model training on minority classes.}}
  \label{fig:cifar10_dp_skew}
  \vspace{-\baselineskip}
\end{figure*}

\begin{table}
  \centering
  \resizebox{\columnwidth}{!}{
  \begin{tabular}{|c|c|c|c|c|c|}
  \hline
  \multirow{2}{*}{Dataset}  & \# Trainable  & Classes               & Train Loss                & Test Loss                 & Non-Private               \\ 
                            & Parameters    &                       & w/ DP (\%)                & w/ DP (\%)                & Vulnerability (\%)        \\
  \hline
  MNIST     & 71,010                        & 10                    & 5.09                      & 3.51                      & 53.11                     \\
  CIFAR-10  & 83,978                        & 10                    & 55.90                     & 14.34                     & 72.58                     \\
  CIFAR-100 & 107,108                       & 100                   & 82.66                     & 49.34                     & 74.04                     \\
  CIFAR-10  & \multirow{2}{*}{1,036,810}    & \multirow{2}{*}{10}   & \multirow{2}{*}{51.00}    & \multirow{2}{*}{26.59}    & \multirow{2}{*}{72.94}    \\
  (with TL) &                               &                       &                           &                           &                           \\
  CIFAR-100 & \multirow{2}{*}{1,071,460}    & \multirow{2}{*}{100}  & \multirow{2}{*}{85.60}    & \multirow{2}{*}{58.39}    & \multirow{2}{*}{89.08}    \\
  (with TL) &                               &                       &                           &                           &                           \\
  \hline
  \end{tabular}}
  \caption{\small{Problem complexity vs membership inference vulnerability and differential privacy utility loss ($\epsilon,\delta)=(10,10^{-5})$).}}
  \label{tab:dputil_vs_mivulnerable}
\end{table}

In Table~\ref{tab:dputil_vs_mivulnerable} we compare the five scenarios (rows) with respect to utility loss and vulnerability of membership inference attacks. The results demonstrate that there is an unfortunate trade-off between the vulnerability and the usability of differential privacy as a mitigation technique. More complex models and classification problems display more significant utility loss when differential privacy is employed during the DNN model training. The MNIST dataset represents a less complex dataset than the CIFAR-10 and CIFAR-100 datasets. In MNIST the images are black and white with MNIST images of the same class being more similar than the colored object images in the CIFAR datasets. Therefore, the MNIST dataset is more easily learned by a small DNN model architecture with reasonable accuracy in the presence of a differentially private optimizer which adds noise perturbation into each training epoch. Not surprisingly, for the same reason, the MNIST model has reduced vulnerability to membership inference attacks as the training examples are more similar to one another.

The CIFAR-10 dataset, on the other hand, is more complex and require more complex models with a greater number of trainable parameters. In this case the models demonstrate a higher vulnerability to membership inference attack than the MNIST dataset. Unfortunately, this complexity also increases the loss in utility when a differentially private optimizer is employed in the model training. 

Additionally, comparing the CIFAR-10 dataset with the CIFAR-100 dataset, CIFAR-100 represents a more complex learning problem with 100 possible classes rather than 10. This again results in not only a more significant loss for the CIFAR-100 dataset in the presence of differentially private deep learning, but also higher vulnerability to membership inference attacks. 

Compare the two models trained on the most complex dataset CIFAR-100, we observe the impact of the model complexity on the vulnerability of membership inference attack. The CIFAR-100 with transfer learning (last row) will have a (24, 24, 3) image input compared to the 50 feature input from the PCA approach using CIFAR-100 (third row). After processing the image through the fixed convolutional layers, the model with transfer learning still has significantly more parameters (compare row 5 with row 3). We note the more complex model structure (over 1 million parameters) has increased membership inference vulnerability for the CIFAR-100 dataset with 89.08\% attack accuracy compared with the vulnerability of 74.04\% when using the less complex model with reduced dimensionality of input (with 107,108 trainable parameters). However the mode complex model also reports a greater test loss of 58.39\% compared to the loss of 49.34\% for the less complex model.

Our experimental results from Table~\ref{tab:dputil_vs_mivulnerable} underscore an important observation. Using differential privacy as effective mitigation for membership inference remains an open challenge. One primary reason is that even with differentially private deep learning, the most vulnerable DNN models (and datasets) are unfortunately also those which experience the greatest utility loss.

\subsubsection{{\bf Implications for Skewed Datasets}}\hfill\\
\vspace{-\baselineskip}

An additional challenge in the deployment of differential privacy is the impact on skewed datasets. Using the approach and model structure from~\cite{jayaraman2019evaluating}, we experiment with different levels of data skewness in the CIFAR-10 dataset. We again gradually decrease the representation of the automobile class from 10\% (even distribution) to 2\% of the training data for a fixed $\epsilon$ of 1000. While this is a large $\epsilon$ value when considering the differential privacy theory, the results in~\cite{jayaraman2019evaluating} indicate that more common theoretical $\epsilon$ values, such as $1$, result in no meaningful learning at all. Our experiments also validate this observation. We conduct this set of experiments by varying the data skewness in the presence of the largest $\epsilon$ reported in~\cite{jayaraman2019evaluating}. In addition, we also measure and report the results for smaller $\epsilon$ with the automobile class set to $2\%$ of its training data. Figure~\ref{fig:cifar10_dp_skew} shows the results of these two sets of experiments, measured using overall model accuracy loss and automobile F-1 score loss. We compare the differentially private deep learning setting with the non-private setting using the same model architecture and data distribution. We report the percentage of utility loss according to these scores. 

In statistical analysis of binary classification, the {\bf F-1 score} measures the accuracy by considering both the precision and the recall of the test to compute the score. The F-1 score is the harmonic mean of the precision and recall. An F-1 score reaches its best value at 1 (perfect precision and recall) and worst at 0.
Let $p$ be the number of correct positive results divided by the number of all positive results returned by the classifier, and $r$ be the number of correct positive results divided by the number of all relevant samples (all samples that should have been identified as positive). We have
$F-1=2(p\times r)/(p+r)$. To isolate the utility loss for the automobile class we therefore report the loss in F-1 score calculated using the precision and recall of only the automobile class where automobile is considered positive and all other classes negative.

On the left sub-figure of Figure~\ref{fig:cifar10_dp_skew}, we compare the accuracy loss of the overall model with the loss in F-1 score of the automobile class (class $1$). We vary the data skewness of the automobile class in the training set, and report the results by the percentage of decrease in F-1 score when using differentially private deep learning setting compared to the non-private deep learning setting. This left sub-figure shows that as the skewness decreases from 2\% to 10\% (X-axis), the gap between the private setting and the non-private setting becomes closer. When the data skewness is high for the automobile class (class $1$), the minority group is more vulnerable, as seen in Figure~\ref{fig:cifar10_distrib}, but unfortunately also shows the greatest utility loss for both training and test when differential privacy is introduced.

On the right sub-figure of Figure~\ref{fig:cifar10_dp_skew}, we again compare the accuracy loss of the overall model with the loss in F-1 score of the automobile class (class $1$). This time we vary the privacy parameter $\epsilon$ from 10 to 1000, and report the measurement results by the percentage of utility loss when using a differentially private deep learning setting compared to a non-private deep learning setting. We observe that the minority group has much higher utility loss under the privacy setting, even when the privacy budget parameter $\epsilon$ is set to larger (and theoretically almost meaningless) values. This confirms the observation that differential privacy as a mitigation technique presents a catch-22 as the minority class is more likely to be successfully targeted by a membership inference attack but also suffers the most under differential privacy. 

In summary, Figure~\ref{fig:cifar10_dp_skew} highlights how differential privacy as a mitigation technique also leads to unfortunate outcomes for minority classes. That is, the data which is most vulnerable to attack is also the most likely to experience intolerable accuracy loss under differentially private setting compared to non-private setting. This mirrors the challenge wherein complex datasets and complex model architectures are more vulnerable to membership inference attacks, and at the same time, are also more likely to experience significant accuracy loss with differentially private deep learning.

\section{Implications for Federated Learning Systems}

In response to legislative restrictions on data sharing, federated learning has gained increased attention from both academics and industry. In a federated learning environment, multiple institutions (or individuals) may leverage their collective data by sharing model parameters rather than sharing raw data. For example, rather than a group of financial institutions sharing their data with one central authority for model training, each can learn a federated model by sharing model parameters and keeping their training data private. Such federated model should be more accurate in prediction than any single participant's model. There are a number of ways one can build such a federated learning model.
First, one can share the local training parameters during each epoch or each iteration of the local model training in a federated learning system. This allows cross-model learning in an iterative manner. Alternatively, one can also build a federated learning model by sharing the model parameters or the locally trained model after each participant has completed its local training over its private training dataset. In this scenario, the federated learning model can be viewed as an ensemble of its participants’ trained models over a horizontally partitioned dataset, which is the union of its participants’ training datasets. 


Recent work on membership inference has shown that federated learning exposes participants to significant privacy risk~\cite{truex2019demystifying, nasr2018comprehensive, wang2019beyond}. In response to such membership privacy vulnerability, proposals for privacy-preserving federated learning have been put forward, including adding differentially private noise prior to the parameter updates by each party~\cite{truex2018hybrid, bonawitz2017practical, zhao2019privacy}. However, preliminary results have indicated that each individual participant is likely to have different privacy risks~\cite{truex2018hybrid}. Also when taking into account the privacy policy of the participating institution, the sensitivity of the data held by that institution, and the distribution of their data, one institution may demand a higher degree of privacy than another participant. We argue that the membership privacy analysis and the membership inference mitigation through differential privacy can be a valuable component in the research and development of federated machine learning systems for real-world mission critical applications.

\section{Related Work}

Existing related research can be broadly classified into three categories: (1) membership inference, (2) the connection between membership inference and adversarial machine learning, and (3) differential privacy as a mitigation technique.

\textbf{Membership Inference Attacks}. The membership inference attack against machine learning models was first presented in~\cite{shokri2017membership} where the authors proposed the shadow model attack. ~\cite{truex2019demystifying} characterized attack vulnerability with respect to different model types and datasets and introduced the vulnerability of loosely federated systems while the authors in~\cite{salem2018ml} relax adversarial assumptions and generate a model and data independent attacker. Hayes et. al. demonstrate the membership inference vulnerability of generative models~\cite{hayes2019logan} while Melis et al. highlight the potential for feature leakage in collaborative learning environments~\cite{melis2018exploiting}. Finally in~\cite{sablayrolles2019white} the authors articulate the power of black-box attacks compared with white-box attacks. We expand on the foundation from these works to offer new insights for the membership inference attack particularly focusing on skewed datasets and differentially private mitigation techniques.

\textbf{Membership Inference and Adversarial Machine Learning}. There have been some recent works in analyzing the connection between membership inference and adversarial machine learning. Song et al.~\cite{song2019privacy} and Mejia et. al.~\cite{mejia2019robust} demonstrate that existing adversarial defense methods cause an increase in membership inference vulnerability. Our membership inference evaluation system highlights the orthogonal risk. That is, how membership inference can inform adversarial machine learning attackers.

\textbf{Differential Privacy to Mitigate Membership Inference}. Both~\cite{jayaraman2019evaluating} and~\cite{rahman2018membership} investigate differential privacy as a mitigation technique for membership inference attacks. Both indicate that existing differential privacy techniques do not display viable accuracy and privacy protection trade-offs. That is, either models show large accuracy losses or display high membership inference vulnerability. We extend this line of work to investigate the implications of DNN model complexity, learning task complexity, and data skewness on membership inference vulnerability and on the effectiveness of differentially private learning as a mitigation strategy.

\section{Conclusion}

Membership inference attacks seek to infer membership of individual training instances of a privately trained model through black-box access to the prediction API of a MLaaS provider. This paper provides an in-depth characterization of membership inference vulnerability from multiple perspectives. We develop MPLens to expose membership inference vulnerability and perform privacy analysis and privacy compliance evaluation. We demonstrate how membership inference attack methods can be used in the development of representative datasets which not only signify a privacy leakage but also facilitate the generation of substitute models with respect to the attack target model, therefore heightening the risk for adversarial attacks. We investigate training data skewness and its impact on membership inference vulnerability. We also evaluate differential privacy as a mitigation technique for membership inference against deep learning models. Our empirical results show that (1) minority groups within skewed datasets display increased risk for membership inference and (2) differential privacy presents many trade-offs as a mitigation technique to membership inference risk. 

\subsection*{Acknowledgement} The authors are partially sponsored by NSF CISE SaTC (1564097) and the first author acknowledges the IBM PhD Fellowship award received in Sept 2019. Any opinions, findings, and conclusions or recommendations expressed in this material are those of the author(s) and do not necessarily reflect the views of the National Science Foundation or other funding agencies and companies mentioned above.

\addtolength{\textheight}{-4cm}  

\bibliographystyle{unsrt}
\bibliography{membership-bib}

\end{document}